# Medium Access Control for Wireless Sensor Networks based on Impulse Radio Ultra Wideband

BERTHE Abdoulaye[1,2], LECOINTRE Aubin[1,2], DRAGOMIRESCU Daniela[1,2], PLANA Robert[1,2]

[1] CNRS; LAAS; 7 avenue du colonel Roche, F-31077 Toulouse, France
[2] University of Toulouse; UPS, INSA, INP, ISAE; LAAS; F-31000 Toulouse, France
{aberthe, alecoint, daniela, plana}@laas.fr

*Abstract.* This paper describes a detailed performance evaluation of distributed Medium Access Control (MAC) protocols for Wireless Sensor Networks based on Impulse Radio Ultra Wideband (IR-UWB) Physical layer (PHY). Two main classes of Medium Access Control protocol have been considered: Slotted and UnSlotted with reliability. The reliability is based on Automatic Repeat ReQuest (ARQ). The performance evaluation is performed using a complete Wireless Sensor Networks (WSN) simulator built on the Global Mobile Information System Simulator (GloMoSim). The optimal operating parameters are first discussed for IR-UWB in terms of slot size, retransmission delay and the number of retransmission, then a comparison between IR-UWB and other transmission techniques in terms of reliability latency and power efficiency.

## INTRODUCTION

Wireless Sensor Networks is a system composed of several autonomous nodes. The nodes' architecture includes a microprocessor, several sensor and actuator devices and also a radio communication device [1-3]. Through their radio communication module, the nodes have the ability to perform wireless communication in order to accomplish a collaborative task. WSNs support a large range of application: monitoring, local area control, factory and house automation, and also tactical applications [1-3].

The intrinsic constraints when dealing with WSNs are reliability, power efficiency, latency efficiency and simplicity. The use of IR-UWB at the PHY layer has been envisioned for WSNs because of its good behaviours which are low radiated power, simple circuitry, high multipath resolution and its resistance against fading [4]. Due to this PHY layer advantages and the nature of the targeted application, the envisioned MAC protocols for IR-UWB have been proven to be ALOHA-like [5].

A key issue when designing a new system is an accurate performance prediction regarding some specific constraints. Thus, a WSN designer may need a performance evaluation tool built on accurate simulation architecture to predict WSNs performances, before the real deployment. However, little has been done in MAC protocols detailed performance evaluation and optimization regarding the previously mentioned WSNs constraints for IR-UWB. The purpose of this paper is to present an accurate and detailed WSN performance evaluation with ALOHA-like MAC protocols. The reliability is considered using the Automatic Repeat ReQuest (ARQ) mechanism [2]. The optimal number of retransmission is first determined for the Slotted and UnSlotted protocols. Then, the optimal retransmission delay and the optimal slot size are presented for the UnSlotted and Slotted protocols. Finally IR-UWB is compared to other techniques in terms of reliability, latency and power efficiency. This paper will be organized as follows: Section 1 presents related work, and then Section 2 presents the modeled MAC protocols for IR-UWB. In

Section 3, the performance evaluation is presented and finally Section 4 concludes.

## 1. RELATED WORK

Modeling and simulation are the commonly used methods to evaluate WSNs performances. As alternative PHY and MAC solutions for the IEEE 802.15.4a standard, several IR-UWB MAC-PHY models have been proposed. In [6], an analytical modeling of pure and slotted ALOHA protocols is presented and a new ALOHA-like protocol is also proposed (UWB)$^2$. Although the analytical study is a useful method to predict performances, it lacks of accuracy compare to a detailed simulation model [7]. In [5], the benefits of using ALOHA-like protocols are demonstrated and based on this, a new MAC protocol: Dynamic Channel Coding (DCC) MAC and a simulation model are proposed. In [8] multichannel distributed protocols: M-ALOHA, M-PSMA, BSMA and their performances evaluation are presented. In [9] an analytical modeling and also a performance evaluation of slotted ALOHA over IR-UWB using an existing simulation platform is presented.

None of these works address WSNs performances optimization regarding its intrinsic constraints (reliability, latency and power efficiency). They do not analyze the impact of the number of retransmission nor the slot size on the system reliability. In this work, we propose a detailed performance evaluation of WSNs; including sensor, sensing channel and a remote detection, and identification application models.

## 2. MAC PROTOCOL FOR IR-UWB

The benefits of ALOHA like Medium Access protocol have been demonstrated for IR-UWB [4][6-10]. Because of the PHY layer characteristics and also the requirements of simplicity. We modeled UnSlotted and Slotted protocols on the Network Simulator GloMoSim. These are lightweight medium sharing protocols. A node immediately transmits once it has a packet to transmit without caring about the channel state (Clear Channel Assessment) CCA mode 3. As specified in the 802.15.4 draft, the channel is always reported to be free with TH-IR-UWB.

In the UnSlotted MAC protocol, once a node has a packet to transmit, it transmits it on the medium. After performing a transmission, the MAC scheduler waits for the acknowledgement of the transmitted packet for a defined duration. When the acknowledgement is received before the expiration of this delay, it transmits the next packet. Otherwise, the transmitted packet must be retransmitted until the number of retransmission exceeds the retransmission limit. The impact of the number of retransmission and the retransmission limit on the reliability and the latency efficiency are presented in the Section 3.

In the Slotted protocol, the medium is divided into time slots and a node is allowed to transmit only at the beginning of a new time slot. This reduces the collision probability as it effectively reduces the number of concurrent transmission being performed. The main advantage of the Slotted protocol is the power management easiness. Once synchronized, the nodes can collaborate and schedule a wake up time to check if there is a packet addressed to them on the medium, which is not possible for the UnSlotted MAC protocol.

## 3. PERFORMANCE EVALUATION

In this section, the simulation results are presented. The simulation scenario consists of a typical WSN application (Figure 1). The sensor nodes are in charge to detect and identify an event of interest and relay it to a base station. This can be used, for instance, in a local area protection or a secured transportation system. The relevant simulation parameters are summarized in Table 1.

The simulations are run on an enhanced GloMoSim (Release 2.02) simulator. The simulated scenario consists of *60 sensor nodes* including the base station, scattered in *140 × 70 m$^2$*. Six targeted nodes represented in the simulation by mobile nodes generate the targeted event at the period of one second with an additional Jitter of 0.1 second. Two of them are equipped with a radio communication interface that allows them to respond to the identification requests sent by the sensor nodes. The used multi hop routing protocol is Ad hoc On Demand Distance Vector (AODV).

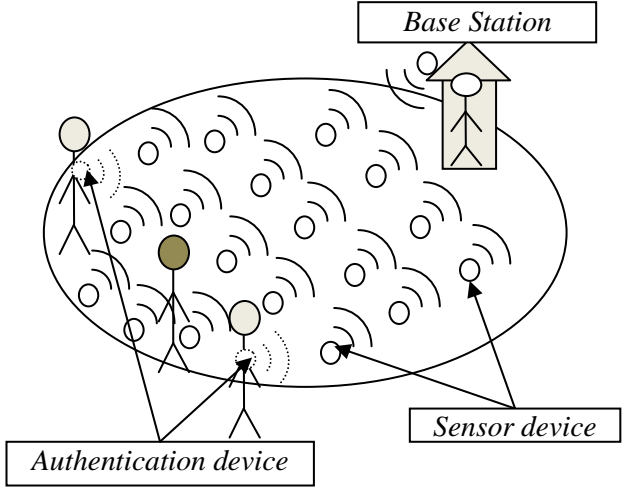

**Figure 1: Simulation scenario**

**Table 1: Simulation parameters**

| Parameter | TH-IR-UWB | OQPSK |
|---|---|---|
| Bandwidth (MHz) | 100 | 2 |
| Frequency (GHz) | 0.8 | 2.45 |
| Throughput (Mbps) | 1 | 0.25 |
| Capture Model | ber based | ber based |
| Antenna Height (m) | 0.45 | 0.03 |
| Antenna Gain (dB) | 3 | 3 |
| Noise Figure (dB) | 5 | 10 |
| Temperature (K) | 270 | 270 |
| Sensitivity (dBm) | -85 | -96 |
| RX-Threshold (dBm) | -80 | -85 |
| TX-Power (dBm) | -24.318 | 17 |

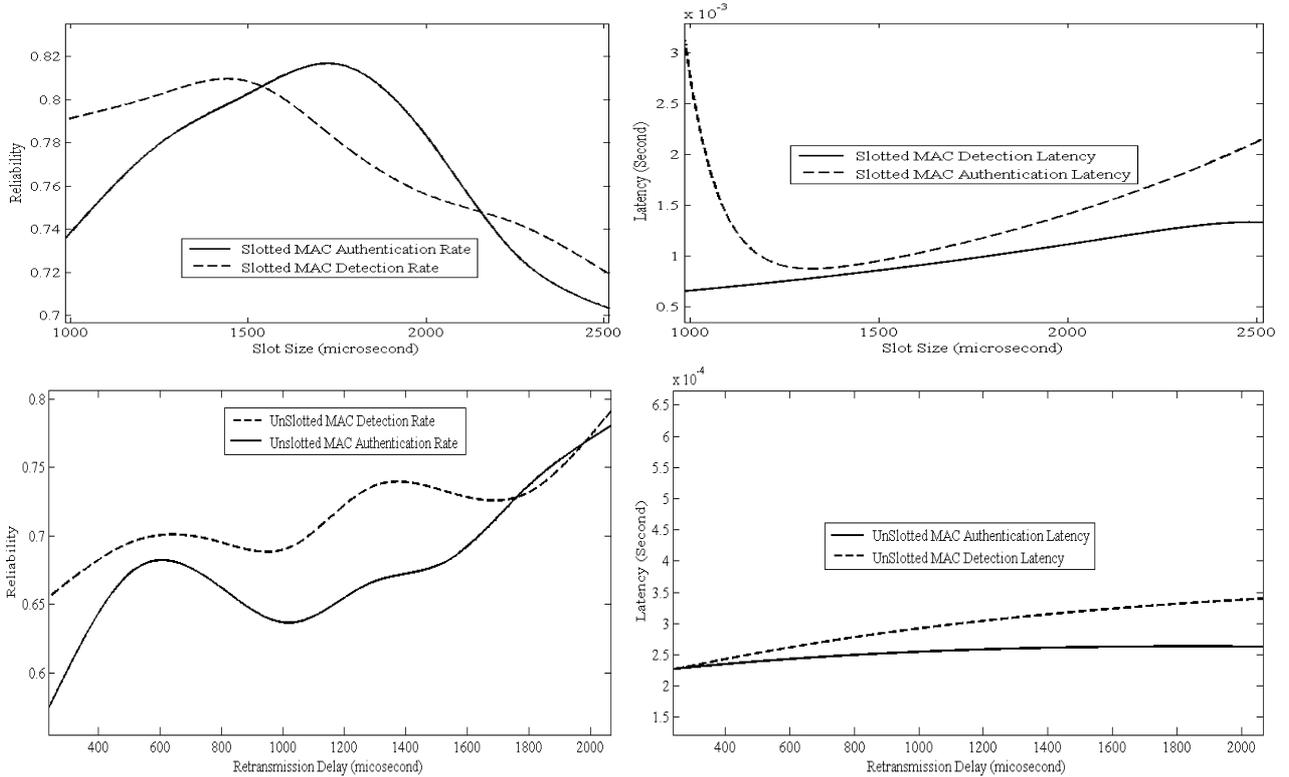

**Figure2: Optimum Operating Parameters: Slot Size and Retransmission Delay**

## Comparative analysis

In Figure 2, we can note that increasing the slot size in the slotted protocol decreases the reliability and leads to high latency. In the UnSlotted protocol, increasing the retransmission delay leads to high reliability, however it leads to high latency. In Figure 3 below, the impact of the retransmission delay is presented for both the Slotted and UnSlotted protocols. It can be seen that up to six retransmissions, increasing the number of retransmission also increases the reliability.

However, when the number of retransmission exceeds 12, it has a negative impact on the reliability and the latency efficiency. We should also note that the Slotted MAC is more reliable than the UnSlotted one but the latter is more latency efficient. For the same scenario, the reliability and the latency efficiency of CSMA over OQPSK was 0.5 and $7.10^{-3}$. Table 2 summarizes the comparison between CSMA over OQPSK and IR-UWB regarding the power consumptions for one day.

**Table 2: Power Consumption**

| Parameter | TH-IR-UWB | OQPSK |
|---|---|---|
| Power consumption (mW-h) | 468 | 1396,8 |
| TX Power Consumption | 5,0 mW/s @ 1.2 V | 52,2 mW/s @ 3V |
| RX Power Consumption | 20 mW/s @ 1.2 V | 59,1 mW/s @ 3V |

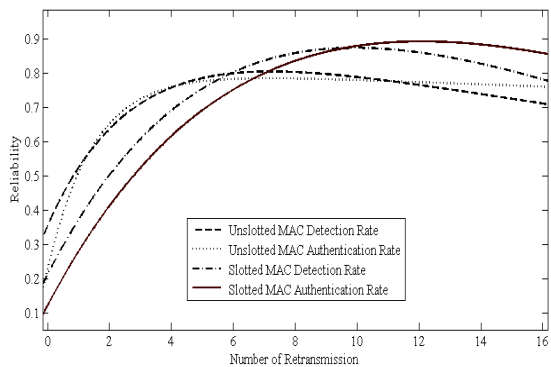
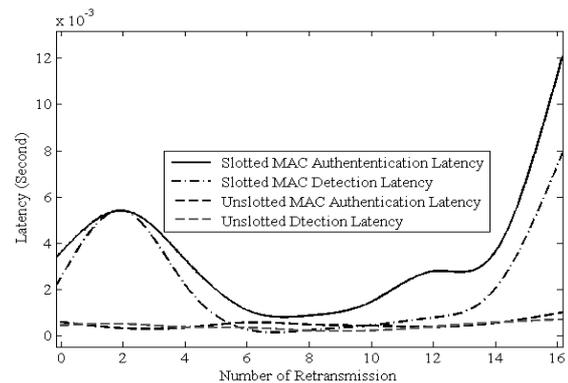

**Figure3: Optimum Operating Parameters: Number of Retransmission**

## CONCLUSION

In this paper we presented Slotted and UnSlotted ALOHA-like MAC protocols performance evaluation for WSN based on IR-UWB regarding the WSN constraints. The optimal operating parameters for IR-UWB have been analyzed in terms of slot size, retransmission delay and retransmission limit. A comparison between CSMA over OQPSK and distributed MAC over IR-UWB has been also presented in the context of low cost, low power and high reliable WSNs.

Our future work will include tracking algorithm performance evaluation and also reconfigurable IR-UWB MAC and PHY implementation.